\documentstyle [epsfig,epsf,rotate]{mn}

\renewcommand{\sun}{\hbox{$_\odot$}}

\newif\ifAMStwofonts
\AMStwofontstrue

\ifoldfss
  \ifCUPmtlplainloaded \else
    \NewTextAlphabet{textbfit} {cmbxti10} {}
    \NewTextAlphabet{textbfss} {cmssbx10} {}
    \NewMathAlphabet{mathbfit} {cmbxti10} {} 
    \NewMathAlphabet{mathbfss} {cmssbx10} {} 
  \fi
  \ifAMStwofonts
    \ifCUPmtlplainloaded \else
      \NewSymbolFont{upmath} {eurm10}
      \NewSymbolFont{AMSa} {msam10}
      \NewMathSymbol{\uDelta}  {0}{upmath}{1}
      \NewMathSymbol{\ubeta}   {0}{upmath}{C}
      \NewMathSymbol{\udelta}  {0}{upmath}{E}
      \NewMathSymbol{\upi}     {0}{upmath}{19}
      \NewMathSymbol{\umu}     {0}{upmath}{16}
      \NewMathSymbol{\upartial}{0}{upmath}{40}
      \NewMathSymbol{\leqslant}{3}{AMSa}{36}
      \NewMathSymbol{\geqslant}{3}{AMSa}{3E}

    \fi
  \fi
\fi 

\ifnfssone
  \newmathalphabet{\mathit}
  \addtoversion{normal}{\mathit}{cmr}{m}{it}
  \addtoversion{bold}{\mathit}{cmr}{bx}{it}
  \newmathalphabet{\mathbfit} 
  \addtoversion{normal}{\mathbfit}{cmr}{bx}{it}
  \addtoversion{bold}{\mathbfit}{cmr}{bx}{it}
  \newmathalphabet{\mathbfss} 
  \addtoversion{normal}{\mathbfss}{cmss}{bx}{n}
  \addtoversion{bold}{\mathbfss}{cmss}{bx}{n}
  \ifAMStwofonts
    \ifCUPmtlplainloaded \else
      \UseAMStwoboldmath
      \makeatletter
      \new@mathgroup\upmath@group
      \define@mathgroup\mv@normal\upmath@group{eur}{m}{n}
      \define@mathgroup\mv@bold\upmath@group{eur}{b}{n}
      \edef\UPM{\hexnumber\upmath@group}
      \new@mathgroup\amsa@group
      \define@mathgroup\mv@normal\amsa@group{msa}{m}{n}
      \define@mathgroup\mv@bold\amsa@group{msa}{m}{n}
      \edef\AMSa{\hexnumber\amsa@group}
      \makeatother
      \mathchardef\uDelta="0\UPM1
      \mathchardef\ubeta="0\UPMC
      \mathchardef\udelta="0\UPME
      \mathchardef\upi="0\UPM19
      \mathchardef\umu="0\UPM16
      \mathchardef\upartial="0\UPM40
      \mathchardef\leqslant="3\AMSa36
      \mathchardef\geqslant="3\AMSa3E
    \fi
  \fi
\fi 

\ifnfsstwo
  \DeclareMathAlphabet{\mathbfit}{OT1}{cmr}{bx}{it}
  \SetMathAlphabet\mathbfit{bold}{OT1}{cmr}{bx}{it}
  \DeclareMathAlphabet{\mathbfss}{OT1}{cmss}{bx}{n}
  \SetMathAlphabet\mathbfss{bold}{OT1}{cmss}{bx}{n}
  \ifAMStwofonts
    \ifCUPmtlplainloaded \else
      \DeclareSymbolFont{UPM}{U}{eur}{m}{n}
      \SetSymbolFont{UPM}{bold}{U}{eur}{b}{n}
      \DeclareSymbolFont{AMSa}{U}{msa}{m}{n}
      \DeclareMathSymbol{\uDelta}{0}{UPM}{"1}
      \DeclareMathSymbol{\ubeta}{0}{UPM}{"C}
      \DeclareMathSymbol{\udelta}{0}{UPM}{"E}
      \DeclareMathSymbol{\upi}{0}{UPM}{"19}
      \DeclareMathSymbol{\umu}{0}{UPM}{"16}
      \DeclareMathSymbol{\upartial}{0}{UPM}{"40}
      \DeclareMathSymbol{\leqslant}{3}{AMSa}{"36}
      \DeclareMathSymbol{\geqslant}{3}{AMSa}{"3E}
    \fi
  \fi
\fi 

\ifCUPmtlplainloaded \else
  \ifAMStwofonts \else 
    \def\uDelta{\Delta}
    \def\ubeta{\beta}
    \def\udelta{\delta}
    \def\upi{\pi}
    \def\umu{\mu}
    \def\upartial{\partial}
  \fi
\fi

\begin{document}

\title[On the precession of accretion discs in X-ray binaries...]
{On the precession of accretion discs in X-ray binaries}

\author[J. Larwood]
{J. Larwood$^{1,2}$  \\ 
$^1$ Issac Newton Institute for Mathematical Sciences,
20 Clarkson Road, Cambridge CB3 0EH \\
$^2$ Astronomy Unit, School of Mathematical Sciences, Queen Mary \&
Westfield College, Mile End Road, London E1 4NS}


\volume{000}
\pagerange{\pageref{firstpage}--\pageref{lastpage}}
\pubyear{0000}

\maketitle

\label{firstpage}

\begin{abstract}

In this \emph{letter} recent results on the nodal precession of
accretion discs in close binaries are applied to the discs
in some X-ray binary systems. The ratio between the tidally forced
precession period and the binary orbital period is given, as well as the
condition required for the rigid precession of gaseous Keplerian
discs. Hence the minimum precessional period that may be supported
by a fluid Keplerian disc is determined. It is concluded that
near rigid body precession of tilted accretion discs can occur
and generally reproduce observationally inferred precession periods,
for reasonable system parameters. In particular long periods in SS433,
Her X-1, LMC X-4 and SMC X-1 can be fit by the tidal model.
It is also found that the precession period that has been tentatively
put forward for Cyg X-2 cannot be accomodated by a tidally precessing
disc model for any realistic choice of system parameters.

\end{abstract}


\section{Introduction}

Some X-ray binary systems show periodic behaviour in their light
curves with periods that are much longer than the binary orbital
period. In order to reproduce the long period evident in the light
curve of Her X-1, Katz (1973) employed a model consisting of an accretion
disc that is tilted with respect to the binary orbital plane and
which is in a state of tidally forced nodal precession. In this
model Katz assumed that only a narrow ring at the disc edge was
tilted and precessing. Gerend \& Boynton (1976) performed detailed
modelling of the observational characteristics of Her X-1
by using a geometrical tilted disc model in which the disc acts as a
shadowing/occulting body and as a source of optical radiation.
Similarly a geometrical model has been applied to SS433 by
Leibowitz (1984), who also related the problem of a precessing
disc to the propagation of tidally excited bending waves.
Heemskerk \& van Paradijs (1989) used a geometrical tilted precessing
disc model to reproduce the light curve of LMC X-4, the third X-ray binary
system with a well determined long period.

Although in these models a rigidly precessing (nodally regressing)
disc was found to produce light curve variations in good aggreement
with the observations, the issue of how a differentially rotating
disc could avoid disruption through differential precession was not
resolved. Papaloizou \& Terquem (1995) showed that rigid body
precession of a Keplerian disc is possible provided that the
sound crossing timescale in the disc is small enough in comparison
with the precession timescale of the disc as a whole. This analytically
derived result was varified numerically by Larwood et al. (1996)
through three-dimensional SPH simulations of warped precessing discs
in close binaries. In that work the precession frequency, derived from linear
perturbation theory for thin discs, was compared with results from the
simulations. The analytical and numerical results were found to be in
good agreement. Here we consider these results in the context of
X-ray binary systems.

In Section $2$ the analytical results are summarised and developed.
In Section $3$ these results are compared with observationally
inferred precession periods for some X-ray binaries, and in Section $4$
discussion is given. In Section $5$ conclusions are presented.

\section{Disc precession}

\subsection{Basic results}

The rigid body precession frequency for a differentially rotating
fluid disc can be derived from linear perturbation theory. For the
leading order term in the part of the secular potential that has
odd symmetry with respect to reflection in the disc midplane,
the induced precession frequency $\omega_{\rm p}$ is given
by (Papaloizou \& Terquem 1995):

\begin{equation}
\omega_{\rm p}=
-\frac{3}{4} \frac{{\rm G} M_{\rm s}}{a^3}
\frac{\int\, \Sigma r^3\, {\rm d}r}{\int\, \Sigma\Omega r^3 \, {\rm d}r}
\cos\delta .
\end{equation}

\noindent $\Sigma (r)$ and $\Omega (r)$ are respectively the
unperturbed axisymmetric surface density and Keplerian angular velocity
profiles. The mass of the secondary (donor) star is $M_{\rm s}$,
the orbital separation is $a$, and $\delta$ is the orbital inclination.
The integrals in this equation are to be taken between the inner
radius $R_{\rm i}$ and the outer radius, $R_{\rm o}$, of the disc.
For a polytropic equation of state with ratio of specific heats
$5/3$ the rigid body precession frequency is readily
calculated (Larwood 1997):

\begin{equation}
\frac{\omega_{\rm p}}{\Omega_{\rm o}} =
-\frac{3}{7}\mu \left(\frac{R_{\rm o}}{a}\right)^3 \cos\delta ,
\label{pfreq}
\end{equation}

\noindent the mass ratio is defined by $\mu = M_{\rm s}/M_{\rm p}$,
where $M_{\rm p}$ is the mass of the primary (compact) object.
In this equation a weak radial dependence of the disc aspect ratio has
been ignored (Larwood \& Papaloizou 1997) and it is assumed that
$R_{\rm o} \gg R_{\rm i}$. This equation differs by a factor of $\sim 2$
from the expression for a precessing ring that was first applied by
Katz (1973). The smaller coefficient found here is due to averaging
of the tidal torque over the disc. Physically this corresponds to a
situation in which the torque that is applied at the outer disc is
transmitted over the entire disc by wave transport or
by viscous stresses, or by a combination of these effects.

We shall use $\alpha$ to denote the usual Shakura-Sunyaev dimensionless
kinematic viscosity (Shakura \& Sunyaev 1973), which is applied
to vertically shearing motions in deriving a dispersion relation
for bending disturbances in linear
theory (Papaloizou \& Lin 1995). The constant mean
Mach number that is associated with the disc is denoted by ${\mathcal M}$. 
In order for bending waves to propagate efficiently it is
approximately required that
$\alpha{\mathcal M} < 1$ (Papaloizou \& Pringle 1983), i.e.
the disc must be sufficiently thick. If this condition is
satisfied then bending waves propagate at approximately one third
of the midplane sound speed $c_{\rm s}$ (Papaloizou \& Lin 1995)
and the condition for
rigid body precession to occur (Papaloizou \& Terquem 1995) is that

\begin{equation}
\frac{r\omega_{\rm p}}{\frac{1}{3}c_{\rm s}} < 1 ,
\label{cond}
\end{equation}

\noindent i.e. the disc crossing time for bending waves is short when
compared with the precession timescale. For standard disc models the
condition (\ref{cond}) will hold everywhere in the disc provided that it is
satisfied near the outer edge. If bending disturbances exist in a
diffusive regime then the propagation speed of bending waves
in (\ref{cond}) should be replaced by the velocity associated with
diffusion of the excited warp. An analysis of the dispersion relation for
bending disturbances in the diffusive regime is given by
Papaloizou \& Lin (1995). In this regime the essentially acoustic
propagation of bending disturbances is inhibited by the damping of
vertically shearing motions. If diffusive behaviour becomes important
in tidally warped discs then the result is a partial loss of internal
communication. Whether the disc can precess as a unit will depend on
how much of the disc is in a state of good internal communication,
if this is a significant portion of the outer disc then the low
angular momentum inner portions can be dragged by the outer regions.
This effect occurs as a result of essentially diffusive communication
between sonically-connected regions in the disc (Larwood et al. 1996). 

\subsection{Development}

The ratio between the binary orbital period and the forced precession
period is obtained from the derived precession frequency (\ref{pfreq}):

\begin{equation}
\frac{P}{P_{\rm p}} =
\frac{3}{7} \mu \left(\frac{1}{1+\mu}\right)^{1/2}
\left(\frac{R_{\rm o}}{a}\right)^{3/2} \cos\delta .
\label{pratio1}
\end{equation}

\noindent The relative size of the disc and orbit can be
eliminated by writing the accretion disc size as a
fraction $\beta$ of the Roche radius $R_{\rm R}$ (cf. Warner 1995),
such that $R_{\rm o} = \beta R_{\rm R}$, then equation (\ref{pratio1})
can be rewritten:

\begin{equation}
\frac{P}{P_{\rm p}} =
\frac{3}{7}\beta^{3/2}
\frac{\mu{\mathcal R}^{3/2}\cos\delta}{\left(1+\mu\right)^{1/2}} .
\label{pratio2}
\end{equation}

\noindent Here the analytical approximation to the Roche radius given
by Eggleton (1983) is used. With this approximation the function
${\mathcal R}\equiv R_{\rm R}/a$ is given by:

\begin{equation}
{\mathcal R} ( \mu ) =
\frac{0.49}{0.6+\mu^{2/3}\ln (1+\mu^{-1/3})} .
\label{eff}
\end{equation}

By using (\ref{pfreq}) the condition (\ref{cond}) can be written in
terms of the binary frequency. Hence the minimum precession period that
can be supported by the disc may be determined as a function of
system parameters:

\begin{equation}
\frac{P_{\rm p}}{P} >
3{\mathcal M}\beta^{3/2} ( 1+\mu )^{1/2}
{\mathcal R}^{3/2} .
\label{cond2}
\end{equation}

\noindent In the diffusive regime this is:

\begin{equation}
\frac{P_{\rm p}}{P} >
8\alpha {\mathcal M}^2\beta^{3/2}
( 1+\mu )^{1/2}
{\mathcal R}^{3/2} .
\label{cond3}
\end{equation}

\noindent Notice that even when $\alpha\sim 1$, diffusive transport cannot
support rigid body precession of the disc as effectively as bending
waves in the non-diffusive regime. This is due to a factor
$\sim\alpha{\mathcal M}$ between the speeds associated with bending wave
propagation and diffusion.

\subsubsection{Disc size}

In order to calculate the precession period for a system of small
orbital inclination we require knowledge of the mass ratio and the
disc size $\beta$. Significant variations in $\beta$ give rise to
significant variations in the calculated period ratio, and since
observations poorly constrain the disc size, it is therefore
desirable to express $\beta$ as a function of mass ratio.

As an accretion disc occupies an increasing fraction of the primary's
Roche lobe, the outer disc feels an increasing tidal effect
and particle orbits increasingly deviate from circular owing to
non-tidal effects. Paczy\'nski (1977) calculated the largest periodic
non-intersecting test particle orbit contained within the primary's
Roche lobe for a range of mass ratios. Paczy\'nski argued that the accretion
disc could not exist outside of this orbit, provided that pressure
and viscosity were sufficiently small in the disc.
Papaloizou \& Pringle (1977) analysed the linearised hydrodynamic
equations for azimuthal modes and calculated the tidal torque
acting on the disc. They determined the radius
at which streamlines would intersect under the action of the
component of the tidal torque with two-fold azimuthal symmetry, for
a range of mass ratios.

Paczy\'nski tabulated the dimensions of the largest test particle
orbit for mass ratios: $0.03 < \mu < 30$. For mass ratios
$0.03 < \mu < 2/3$ it is found that the mean disc
radii (hereafter ``Paczy\'nski radii'')
are roughly independent of mass ratio with a mean value of approximately
$\beta = 0.86$. For $2/3 < \mu < 30$ Paczy\'nski's values give
$\beta$ as a decreasing function of mass ratio, with values in the
range $0.85$ -- $0.6$. In this case the dependence of the mean
radius on mass ratio can be fit with the function:

\begin{equation}
\beta_{\rm P} ( \mu ) = \frac{1.4}{1+[\ln ( 1.8\mu )]^{0.24}} .
\label{pacrad}
\end{equation}

\noindent This function fits the Paczy\'nski radii to better than
$1\%$ for $2/3 < \mu \la 15$ and to better than $3\%$ for
$15 \la \mu \la 30$. Hence if hydrodynamical effects can be ignored,
i.e. if the disc is sufficiently thin and weakly viscous,
the disc size can be determined as a function of mass ratio.
This estimate for the disc size is a lower limit for accretion
discs in real systems since even small amounts of gas pressure
or viscosity can result in larger discs (Papaloizou \& Pringle 1977).

Papaloizou \& Pringle tabulated values of the disc size resulting
from the two-fold symmetric component of the tidal torque for
mass ratios: $0.2 < \mu < 10$. It is found that the dependence on
mass ratio for these values is roughly
a constant, having a mean value $\beta\approx 0.88$.
It is noted that this estimate is least reliable for small mass
ratios since higher order azimuthal modes can become significant in
these cases, which results in a smaller disc size.

The actual truncation radii of accretion discs in close binaries
will depend on the hydrodynamic properties of the disc gas. If
the disc is weakly viscous and pressurised then the Paczy\'nski
radii are expected to describe accurately the the disc sizes. However,
if the disc is thick and/or viscous then these values may under
estimate the disc size by a factor of up to $\sim 2$, for mass ratios
significantly greater than unity.

\section{Application}

Data for all the objects discussed here can be found in Table~(1).
\begin{table}
\centering
\begin{minipage}{70mm}
\caption{Data used for X-ray binaries in this paper.
References are given in the text.}
\begin{tabular}{ccccc}
\hline
Source & $M_{\rm s}({\rm M}\sun )$ & $M_{\rm p}({\rm M}\sun )$ &
$P_{\rm p}({\rm d})$ & $P({\rm d})$ \\
\hline
\\
 SS433   &  3.2  & 0.8  & 164  & 13.1 \\
 Her X-1 &  2.3  & 1.5  & 35   &  1.7 \\
 LMC X-4 & 15.7  & 1.48 & 30.5 &  1.4 \\
 Cyg X-$2^{a}$
         &  --   &  --  & 78   &  9.8 \\
 SMC X-1 & 17.2  & 1.6  & 55   &  3.9 \\
\hline
\end{tabular}
${}^{a}$Cyg X-2 has an estimated mass ratio of 0.34.
\end{minipage}
\end{table}
\subsection{Systems with well determined periods}

In this Section the formulae presented above are applied on a
case by case basis to observed systems having well determined
long periods (e.g. White, Nagase \& Parmar 1995).

\subsubsection{SS433}

A review of the observational characteristics of this object has been
given by Margon (1984). SS433 exhibits a bipolar jet with an
anti-symmetric S-curve structure. The opening angle of the S-curve
is approximately $40\degr$. The standard naive picture for this
system involves a precessing disc with the precessing jet being
launched normal to the disc surface and from the central regions of
the disc. Within the context of this interpretation we deduce that
the disc is inclined by $20\degr$ with respect to the orbital plane
of the binary.

The binary system has a mass ratio $\mu = 4$ (D'Odorico et al. 1991)
and the period ratio is determined to be ${P_{\rm p}}/{P}= 13$. Using
these values equation (\ref{pratio2}) implies $\beta = 0.84$, which is
consistent with the truncation radii calculated by
Papaloizou \& Pringle (1977) for gas discs. The condition
(\ref{cond2}) then indicates that in order to support the observed
level of precession we require a Mach number ${\mathcal M} < 18$.
Features of the long period in the photometric light curve
suggest that the disc may be very thick with an aspect ratio
significantly greater than $0.1$ (Margon 1984),
which could permit the quasi-rigid body precession of the disc under
the condition given above. A further consideration is of the short
period photometric variations that are thought to be due
to ``nodding'' motions of the disc (Katz et al. 1982). The
period of these variations is one half of the synodic period of
the binary. It was suggested that the disc must be able to
efficiently communicate with itself in order to give a coherent
photometric variation and jet response to the time
varying tidal torques that are applied to the outer disc.
Qualitatively similar nodding behaviour was found in the
numerical simulations of Larwood et al. (1996), for a disc model
with an aspect ratio $\sim 0.15$.

\subsubsection{Her X-1}

This system has a mass ratio $\mu = 1.5$ (Reynolds et al. 1997)
and a period ratio ${P_{\rm p}}/{P}= 21$. In
order to reproduce the observed period ratio we find approximately
$\beta = 0.7$, from equation (\ref{pratio2}) for small inclination.
This value is the same as the Paczy\'nski radius, being smaller
than the truncation radius expected in a gas disc (Papaloizou \& Pringle 1977).
The condition (\ref{cond2}) indicates that in order to support the
observed level of precession we require a Mach number
${\mathcal M} < 38$. The radiative heating model of
Schandl \& Meyer (1994) gives a Mach number close to this value for
Her X-1. Our result is consistent with the tidal
precession of a thin disc. It is noted that this situation is not in
disagreement with inferences of large aspect ratios for this system
since thin discs can be strongly warped with large
opening angles (Larwood et al. 1996).

However there is independent evidence that the disc in Her X-1 might
be quite thick. Katz et al. (1982) find evidence for nodding motions
in the photometric data, and detailed modelling of the optical
characteristics by Gerend \& Boynton (1976) results in a model fit with
an extended thick disc with a large inclination. Gerend \& Boynton find
$\beta =0.77$ and $\delta = 30\degr$. Remarkably, with these values
equaion (\ref{pratio2}) yields a period ratio of $21$. It is not known
whether a disc with such
a high inclination can survive in a semi-detached system, yet it is
amazing that parameters deduced from a geometrical model for generating
the light curve are consistent with the tidal model for disc precession.

\subsubsection{LMC X-4}

This is an extreme mass ratio system with
$\mu = 10.6$ (Levine et al. 1991) and a period ratio
${P_{\rm p}}/{P}= 22$ (Lang et al. 1981). For a small disc inclination
and the expected Paczy\'nski radius with $\beta = 0.6$, we
find a period ratio ${P_{\rm p}}/{P}\approx 19$. This level of
precession may be supported if ${\mathcal M} < 48$. As for Her X-1,
we deduce a thin precessing disc for LMC X-4. It is noted that
in order to match the observed period exactly we require an
inclination $\delta = 35\degr$, or a disc size $\beta\sim 0.5$.

As for SS433 and Her X-1, LMC X-4 shows evidence for nodding
motions (Heemskerk \& van Paradijs 1989). However if this is
taken as evidence of a much more extended disc then we would
have to accept extreme inclinations (e.g. $\delta\sim 60\degr$
for $\beta = 0.8$).

\subsection{Other systems}

There are few other systems reported in the literature which are
known to exhibit periodic X-ray variability on long timescales
and which also have known orbital periods and for which there
exists an estimate of the mass ratio. Cyg X-2 and SMC X-1 are
two such systems.

\subsubsection{Cyg X-2}

This system has recently been identified to have a period ratio
of ${P_{\rm p}}/{P}= 8$ (Wijnands, Kuulkers \& Smale 1996).
Although the origin of this period has been
tentatively proposed as resulting from disc precession the available
data is at present insufficient to make detailed comparisons with
the systems discussed above (Wijnands et al. 1996).
The binary has a mass ratio
$\mu = 0.34$ (Casares, Charles \& Kuulkers 1998),
which indicates a Paczy\'nski radius with $\beta = 0.86$.
Using these values implies a period ratio of $31$, for small
inclinations. Therefore the observed period ratio for this
system \emph{cannot} be reproduced for any $\beta < 1$,
or any inclination. Also the observed period cannot be obtained by
even quite large adjustments to the mass ratio (see below).
The condition for rigid precession at the observed rate is also
very stringent: ${\mathcal M} < 9$. Assuming an accurate
determination for the orbital period, it would seem very unlikely
that the usual tilted precessing disc model can apply to
Cyg X-2 and another explaination for the long period must
be sought instead.

\subsubsection{SMC X-1}

This system, with mass ratio $\mu = 10.8$ (Reynolds et al. 1993),
is thought to show a period ratio $\sim 14$ (Wojdowski et al. 1998).
For small inclinations we find that
$\beta = 0.7$ is consistent with the observed period, being larger
than the Paczy\'nski radius with $\beta =0.6$. As for SS433 a
thick disc is also required here, with ${\mathcal M} < 25$.
This is consistent with the larger disc size than for LMC X-4,
which has a similar mass ratio. Slightly more extended discs
can be considered for high inclinations (e.g. $\delta = 30\degr$ for
$\beta = 0.77$).

\section{Discussion}

\begin{figure}
\centering{\epsfig{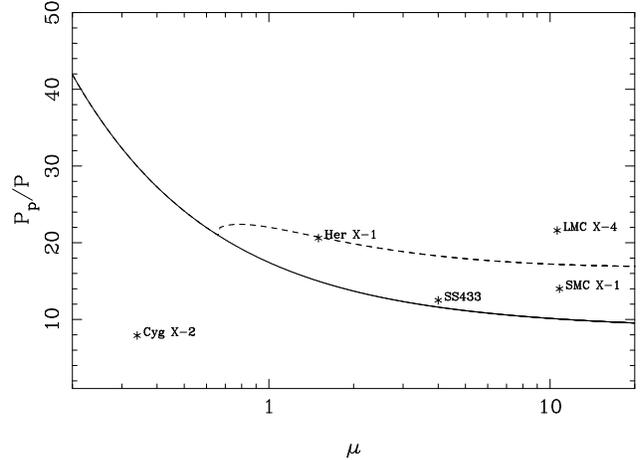}}
\caption{The period ratio against mass ratio for a uniform
truncation radius with $\beta = 0.87$ (solid line) and also
for $\beta = \beta_{\rm P}(\mu)$ (dashed line), the latter
is only used for $\mu > 2/3$. It is assumed that $\cos\delta\sim 1$.}
\label{fig1}
\end{figure}

For mass ratios $\mu = 0.2$ -- $20$, with Paczy\'nski
radii corresponding to values $\beta = 0.9$ -- $0.6$, and
for small inclinations, we find period ratios
for precessing tilted discs in the range
${P_{\rm p}}/{P}\sim 18$ -- $40$. If the mean Papaloizou-Pringle
disc truncation radius applies then the lower limit could be $\sim 10$.
In Fig.~1 we give the period ratio for a constant $\beta =0.87$ for
all mass ratios, which averages the mean constant values given by
the truncation models described earlier. The period ratio for
$\mu > 2/3$ that uses the analytical fit (\ref{pacrad}) to the
Paczy\'nski radii is also plotted, as are the values given in Table~(1).

Given our ignorance of the disc sizes in these systems the period
ratios for SS433, Her X-1, LMC X-4 and SMC X-1 are fit reasonably well
by the tidal model for disc precession. LMC X-4 is fit the worst as it
appears to require an uncomfortably high inclination and/or a very small disc.
However, since the mass ratios for LMC X-4 and SMC X-1 are similar,
the large difference in their period ratios is probably due to differences
in disc size. Therefore the hydrodynamic properties of the discs
should be different, or the disc truncation process should be
different, in the two systems. It is noted that wind accretion
can be important in the mass transfer process in high mass ratio X-ray
binaries (e.g. Frank, King \& Raine 1985). The slow stellar winds that
are thought to exist in LMC X-4 and
SMC X-1 (Hammerschlag-Hensberge, Kallman \& Howarth 1984)
favour the formation of
small discs (e.g. Theuns \& Jorissen 1993), however the
effect of stellar winds on the
truncation of a viscously expanding disc is not known.
The smaller orbital separation in the LMC X-4 system would appear to be
consistent with the idea of wind-driven truncation of the discs.

In the simulations of Larwood et al. (1996) the precession periods
of the simulated discs were found to be $\sim 10\%$ longer than
the expected values when the discs were moderately thin,
i.e. when the level of internal communication of the disc was relatively poor.
It is of interest to note that SS433 appears to be in a state of
good internal communication and we infer a disc size comparable to
that expected for gas discs.
LMC X-4 requires a comparatively small disc, although
larger discs could reproduce the required precession rates
by a modest relaxation of internal communication in the disc.
This is reasonable since the larger period ratio for
LMC X-4 permits a thinner disc than for SS433, and so may
marginally satisfy $\alpha{\mathcal M} > 1$.

Fig. 1 shows that for $\mu >1$ the period ratio is not
a sensitive function of mass ratio. Hence for the systems considered above,
the most important effects for modifying the expected period ratios are
loss of internal communication and changes to the disc size.
It is also noted that the $\la 10\%$ fluctuations that are often seen
in the precession periods of the objects discussed here can be
accomodated by the tidal model by assuming small fluctuations in the disc
size, perhaps due to time dependent tidal torques that cause time
varying distortions to the envelope of the outer disc. In this case
fluctuations in the precession rate should be minimised for the
smallest discs.

\section{Conclusions}

On a case by case basis the analytical results for the tidally induced
precession of tilted accretion discs agree reasonably well with
observationally inferred period ratios for
SS433, Her X-1, LMC X-4 and SMC X-1. The results for SS433, Her X-1
and SMC X-1 are consistent with inferences of thick discs that
occupy a substantial fraction of the Roche lobe radius. This is
encouraging for the current model since thick discs are expected
to allow the propagation of bending waves over much of the disc,
and so enable its quasi-rigid body precession despite Keplerian shear.
It is noted that in the diffusive regime the maximum Mach numbers
quoted above may be smaller by a factor of $\sim\alpha{\mathcal M}$.
If those discs have $\alpha{\mathcal M}$ significantly greater than
unity then we would need extreme disc thicknesses, and therefore
the precessing disc model might become untenable in terms of rapidly
communicating across the disc a precessional torque that is
essentially applied to the outer regions. Unfortunately there
are currently no unambiguous observational constraints on the
disc thickness in any of these systems.

LMC X-4 is not fit so well as the other systems,
but its extreme mass ratio and its short orbital period may
require a different
disc formation mechanism than was supposed here. If wind accretion
becomes important as a mass transfer process then
the disc size is generally smaller than that which results
from standard Roche lobe overflow. If the disc remains small (perhaps
owing to the incident stellar wind from the secondary) then
the period ratio in LMC X-4, and other extreme mass ratio
systems, may be larger than predicted from theory given only
knowledge of the mass ratio. In this case the disc would not
necessarily need to be thin, as is required here to account for
the small disc size that is needed to give the observed precession
rate.

For mass ratios greater than about $2/3$ the disc size is not
well constrained by
theory (we expect $0.9 > \beta > \beta_{\rm P}$). Therefore
in order to make a stronger comparison with the observations
requires further observational input on the sizes of the discs in
these systems.

The precession period that has recently been attributed to Cyg X-2
cannot be due to a simple tidally precessing tilted disc model
because the long period for that system is much too short,
even if the mass ratio is very much larger than currently believed.
It is expected that quasi-rigid body precession of the disc at the
observed rate cannot be supported by the disc whatever the origin of the
precessional torque acting on the outer disc.

In this paper we have not confronted the all important issue of an
origin for the disc tilt. Several mechanisms for either setting up
inclined discs, or allowing for small tilts and warps to grow, have so
far been proposed. In the context of a tidal model,
Papaloizou \& Terquem (1995) deduced that the secular response of the
disc allows for the disc inclination to increase by significant
amounts over a viscous timescale. Radiative torques have also been
put forward as a possible origin for warping and precession in X-ray
binaries (Maloney \& Begelman 1997). This model involves applying the
precessional torque at the outside of a cold disc, and communicating it
throughout the disc by diffusive effects alone. As indicated above,
even if $\alpha\sim 1$ it seems unlikely that very thin discs could survive
precession at the observed rates. This process has yet to be studied
in the non-diffusive regime.

\section*{Acknowledgments}

JL thanks Jonathan Katz, Andrew King, Jean-Pierre Lasota, John Papaloizou,
and Caroline Terquem for valuable discussions and comments. In the
preparation of this paper use was made of STARLINK facilities at
Queen Mary \& Westfield College.



\label{lastpage}
\end{document}